\documentclass[prc,superscriptaddress]{revtex4}
\usepackage{amssymb,color,graphicx,bm}

\definecolor{red}{rgb}{0.8,0,0}
\definecolor{RED}{rgb}{0.8,0,0}
\definecolor{violet}{rgb}{0.4,0,0.4}
\definecolor{green}{rgb}{0,0.5,0.0}
\definecolor{GREEN}{rgb}{0,0.5,0.0}
\definecolor{navy}{rgb}{0.0,0.0,0.6}
\definecolor{orange}{rgb}{0.8,0.2,0.0}
\definecolor{blue}{rgb}{0.3,0.0,0.8}

\begin{document}

\title{Modulus stabilization in a non-flat warped braneworld scenario}

\author{Indrani Banerjee}
\email{indrani.bannerjee@bose.res.in}
\affiliation{Department of Astrophysics \& Cosmology\\ S. N. Bose National Centre for Basic Sciences\\ 
Block JD, Sector III, Salt Lake City, Kolkata-700106}
\author{Soumitra SenGupta}
\email{tpssg@iacs.res.in}
\affiliation{Department of Theoretical Physics\\
Indian Association for the Cultivation of Science\\
2A \& 2B Raja S C Mullick Road, Kolkata-700032, India}

\begin{abstract}
The stability of the modular field in a  warped brane world scenario has been a subject of interest for a long time. Goldberger \& Wise ( GW ) proposed a mechanism to achieve this by invoking a massive scalar field in the bulk space-time neglecting the back-reaction.  In this work, we examine the possibility of stabilizing the modulus without bringing in any external scalar field.
We show that instead of  flat 3-branes as considered in Randall-Sundrum ( RS )  warped  braneworld model, if one considers  a more generalized version of warped geometry with  de-Sitter 3-brane, then 
the brane vacuum energy automatically  leads to   a modulus potential with a metastable minimum. Our result further reveals that 
in this scenario the gauge hierarchy problem can also be resolved for appropriate choice of brane cosmological constant.

\end{abstract}

\maketitle


\section{Introduction} 

After the discovery of the Higgs like scalar at 126 GeV \citep{LHC},  the fine tuning problem of Higgs mass, originating from it's large radiative correction, has became a pressing issue to be resolved within 
the framework of various proposals  beyond standard model Physics.  Among various extra dimensional models \cite{ADD,Antoniadis,Horava,RS,RS2,ADD1,Kaloper,Cohen} the ADD \citep{ADD} and the RS \citep{RS} model drew special attention to
address  this problem . In particular, the RS \citep{RS} model keeps the Higg's mass well within the acceptable upper bound  without invoking any hierarchical parameters in the theory. 
The phenomenology that follows in such a beyond standard model ( BSM ) Physics \cite{DHR, PDBMSSG, TPSSG} is crucially dependent on the stable value of the radius modulus of the model which determines the various parameters in the 4-dimensional effective theory of the model. 
However, the issue of stabilizing the modulus , consistent with the requirement of the solution of gauge hierarchy problem, was proposed by Goldberger \& Wise \citep {GW99} by invoking an additional scalar field in the bulk. By choosing the appropriate boundary values of this scalar at the two orbifold fixed points they could obtain the appropriate value for the warp factor. Subsequently the RS model was generalized with curved 3-branes, in particular with a de-Sitter (DS) and anti de-Sitter (ADS) 3-brane sitting at the orbifold fixed points \citep{DMS}. The phenomenological implications of this have also been discussed in \cite{DMS,JMSSG}.
Here, we demonstrate that for de-Sitter 3-brane, a positive value of the brane vacuum energy  can naturally lead to a modulus potential which has a metastable 
minimum. The value of the corresponding radius modulus depends on the brane cosmological constant and can be tuned to address the gauge hierarchy and stability issue concomitantly. Thus, we do not require any external bulk scalar to achieve modulus stabilization.
In this context it may be mentioned that Csaki et.al \citep{csaki} explored  how the three  brane matters namely matter on either of the two branes or in bulk may contrubute to Hubble expansion on the brane. They showed that in such scenario how  
 the Hubble expansion equation on the brane comes with correct signature only after  assuming a  stabilizing mechanism for the modulus. 
 In our work we demonstrate how  the modulus  stabilization can be achieved in presence of  brane vacuum energy  without introducing any other bulk scalar potential.


\section{A brief survey of warp geometry in non-flat 3-branes}
RS model is formulated in a 5-dimensional anti-de Sitter bulk spacetime where the extra dimension is compactified in a $\rm M^4 \times S^1/Z_2$ manifold.
In the RS model two 3-branes are placed at the two orbifold fixed points which 
are assumed to be flat such that the cosmological constant
induced on the visible brane is zero.  It has been shown that an appropriate  tuning between the brane tension and the bulk induced cosmological constant on the brane can exactly cancel the resulting effective cosmological constant on the brane to make it flat \citep{SSM}.
However a slight imbalance between these may lead to positive or negative brane vacuum energy which leads to warped geometry models with non-flat 3-branes. 

A generalized version of the RS model with non-flat 3-branes was addressed in \citep{DMS}  where 
a more general warp factor was derived and the  correlation between the extra dimensional modulus and brane cosmological constant  was discussed in the light of gauge hierarchy problem.

The metric ansatz in the generalized-RS scenario, satisfying the 5-d Einstein equations with a negative bulk cosmological constant is \citep{DMS}:
   
\begin{equation}
{ds}^2=e^{-2A(r_c \phi)}g_{\mu \nu} {dx}^{\mu}{dx}^{\nu} + r_c ^2 {d\phi}^2.
\end{equation}

This results in the 4-d Einstein equations,

\begin{equation}
^4 G_{\mu \nu} - g_{\mu \nu}e^{-2A}[-6 A'^2 +3 A'']=-\frac{\Lambda}{•4M^3}g_{\mu \nu}e^{-2A},
\end{equation} 
while the component of the equation along the extra dimension is,
\begin{equation}
-\frac{1}{•2}e^{2A} ~ ^4 R+6A'^2=-\frac{\Lambda•}{4M^3•}
\end{equation}
along with the boundary conditions
\begin{equation}
[A'(y)]_i=\frac{\epsilon_i}{24M^3•}V_i.
\end{equation}

Here, $\Lambda$ is the bulk cosmological constant, $R$ the bulk Ricci scalar and $V_i$
the brane tension on the $i^{th}$ brane, where i=vis(hid) for visible (hidden) branes respectively and $\epsilon_{hid}=-\epsilon_{vis}=1$.

On solving equation (2) and equation (3) one obtains the equations:

\begin{equation}
6A'^2=\frac{\Lambda}{4M^3} + 2\Omega e^{2A}
\end{equation}  
and
\begin{equation}
3 A''=\Omega e^{2A}
\end{equation} 
where $\Omega$ represents the effective cosmological constant induced on the 
brane.

Although $\Lambda < 0$ (AdS bulk), $\Omega$ can be both positive and negative ($\Omega<0$ for AdS and $\Omega>0$ for DS spacetimes respectively).\\

\section*{de Sitter Brane}
When $\Omega >0$ the solution of equations (5) and (6) gives the generalized warp factor as:
\begin{equation}
e^{-A}=\omega sinh(ln\frac{c_2}{\omega}-kr_c |\phi|),
\end{equation}  

where, $\omega=\frac{\Omega}{3k^2•}$ and $c_2=1+\sqrt{1+\omega ^2}$.

On equating the ratio of the Higgs mass and the Planck mass to the warp factor at the visible brane,
\begin{equation}
\frac{m}{m_0}=10^{-16}=e^{-A}=\omega sinh(ln\frac{c_2•}{\omega}-kr_c \pi), 
\end{equation}  
one obtains,
\begin{equation}
e^{-kr_c \pi}=\frac{10^{-16}}{c_2}[1 + \sqrt{1+\omega ^2 10^{32}}].
\end{equation}

It is important to note here, that $e^{-kr_c \pi}$ assumes a definite solution for every value of $\omega$. The value of the present day cosmological constant $\Omega \sim 10^{-124}$ (in Planckian units) yields  $kr_c \pi\sim 16ln10$ which in turn produces the required warping of the Higgs mass on the visible brane as in the RS scenario.\\

\section*{Anti-de Sitter Brane}
When $\Omega < 0$ the solution of equations (5) and (6) yields the generalized warp factor  to be:

\begin{equation}
e^{-A}=\omega cosh(ln\frac{\omega}{c_1•}+kr_c |\phi|),
\end{equation}  

where, $\omega=-\frac{\Omega}{3k^2•}$ and $c_1=1+\sqrt{1-\omega ^2}$.

Once again  to solve the gauge hierarchy problem, the ratio of the Higgs mass $m$ to the Planck mass $m_0$ at the visible brane, i.e., $\phi=\pi$ yields,
\begin{equation}
\frac{m}{m_0}=10^{-16}=e^{-A}=\omega cosh(ln\frac{\omega}{c_1•}+kr_c \pi).
\end{equation}  

which implied that,

\begin{equation}
e^{-kr_c \pi}=\frac{10^{-16}}{c_1}[1 \pm \sqrt{1-\omega ^2 10^{32}}].
\end{equation}

This led to $\omega ^2 \le 10^{-32}$, thereby establishing a constrain on the upper limit of $\omega$.
It is quite clear from equation (9) that $e^{-kr_c \pi}$ has two solutions for every allowed value of $\omega$ except for $\omega ^2 = 10^{-32}$, and both the values give rise to the necessary warping.   
\\
\\
Thus the generalized RS scenario \citep{DMS} incorporates the effects of the brane cosmological constant  on the  warp factor and also successfully addresses the gauge hierarchy problem such that the Higgs mass on the visible brane is appropriately warped to the TeV scale. The fact that the cosmological constant can assume such small values and yet address the hierarchy issue implies  that the  gauge hierarchy problem and the cosmological fine tuning problem, are interlinked.

\section{Modulus potential in a curved brane scenario}
In the generalized RS scenario \citep{DMS}, $r_c$ represents the distance between the two branes. However, $r_c$ is assumed to have a stable non-zero value in both the cases. The mechanism of generating this stable value is not addressed in this model. 
Hence, the next step would be to treat 
the distance between the two branes as a 4-dimensional field, the so called radion field or the modular field, $T(x)$, whose vacuum expectation value would be $r_c$. 
An appropriate  mechanism is now required to stabilize the modulus to its vacuum expectation value by generating a a potential term for the modulus in the effective 4-dimensional Lagrangian. 
Goldberger \& Wise \citep{GW99} introduced a bulk scalar field in the RS \cite{RS} Lagrangian which in turn generated a 4-dimensional potential 
for $r_c$. The parameters of this potential are dependent on the newly introduced terms in the action and could be adjusted such that the minima of the potential settles to the value of $r_c$ to generate the required warping. Goldberger \& Wise \citep{GW99}, in their analysis  assumed flat 3-branes as in the original RS \cite{RS} scenario and neglected any possible back-reaction of the stabilizing bulk scalar on the background metric. Motivated by the non-flat warped geometry scenario \citep{DMS} as discussed in the previous section we now explore the possibility of 
stabilizing the modulus $r_c$, without invoking any external scalar field.
In this work, we investigate the stability of the radion field from a modular potential which may be generated due to the non-flat character of the 3-branes at the orbifold boundaries.

The metric ansatz we consider is the following:
 
\begin{equation}
{ds}^2=e^{-2A(x_\mu,\phi)}g_{\mu \nu} {dx}^{\mu}{dx}^{\nu} + T(x)^2 {d\phi}^2
\end{equation}

which satisfies the Einstein's equations obtained from the action:

\begin{equation}
S=S_{gravity}+S_{vis}+S_{hid}
\end{equation}

\begin{equation}
S_{gravity}=\int_{-\infty} ^{\infty} d^4x \int_{-\pi} ^{\pi} d\phi \sqrt{-G}(2M^3 R - \Lambda )
\end{equation}

\begin{equation}
S_{vis}=\int_{-\infty}^{\infty} d^4x \sqrt{-g_{vis}} (L_{vis} -V_{vis})
\end{equation}

\begin{equation}
S_{hid}=\int_{-\infty}^{\infty} d^4x  \sqrt{-g_{hid}}(L_{hid} -V_{hid})
\end{equation}

\subsection{•Case A: de-Sitter three branes}
In this section we consider de-Sitter 3 branes for which the generalized warp factor is given by,

\begin{equation}
e^{-A}=\omega sinh(ln \frac{c_2}{\omega} - k|\phi|T(x)),
\end{equation}

Considering the above form of the warp factor, the first term of equation (15) becomes,

\begin{eqnarray}
\nonumber S_{gravity}^{(1)}=-2M^3\int d^4x \int_{-\pi}^{\pi} d\phi ~ \sqrt{-g} e^{-2A} [T(x)~ {^4}R + 8k^2e^{-2A} T(x) +12k^2e^{-2A} T(x)coth^2 (ln \frac{c_2}{•\omega}-kT(x)|\phi|) \\ \nonumber
+6k|\phi| T(x),^\alpha T(x),_\alpha coth(ln \frac{•c_2}{•\omega} - kT(x)|\phi|)-6k^2|\phi|^2 T(x) T(x),^{\alpha}T(x),_{\alpha}coth^2(ln \frac{•c_2}{•\omega} - kT(x)|\phi|)] 
\end{eqnarray}

while the second term gives,

\begin{equation}
S_{gravity}^{(2)}= -\int d^4 x \int_{-\pi}^{\pi} e^{-4A} \sqrt {-g} ~ \Lambda
\end{equation}

The effective action which is obtained by integrating over the higher dimension is,

\begin{eqnarray}
^4S = \int d^4 x (^4S_{(1)} +^4S_{(2)} +^4S_{(3)})
\end{eqnarray}

where, 

\begin{equation}
^4 S_{(1)}=-2M^3 \int d^4 x \sqrt {-g} ^4R[\frac{c_2 ^2}{4k}+\frac{\omega^2}{k•}ln\frac{\Phi}{f•}+\frac{\omega ^4}{4 k c_2^2•}\frac{f^2}{\Phi ^2•}-\frac{\omega^4}{4kc_2^2•}-\frac{c_2 ^2\Phi ^2}{4kf^2•}]
\end{equation} 
\begin{equation}
^4 S_{(2)}= \int d^4 x \sqrt{-g} [- \frac{1}{2•}\partial_\mu \Phi \partial^\mu  \Phi - 4 \frac{M^3}{k}\omega^2
(\frac{1}{\Phi ^2•}ln\frac{\Phi}{f•})\partial_\mu \Phi \partial^\mu  \Phi + \frac{3M^3}{•k}\frac{\omega^4}{c_2 ^2•}f^2 \frac{•\partial^\mu  \Phi \partial^\mu  \Phi}{•\Phi ^ 4}] 
\end{equation}
\begin{eqnarray}
^4 S_{3}=-2M^3 \int d^4 x \sqrt{-g} [V(\Phi)]
\end{eqnarray}
Here, $\Phi = fe^{-kT(x)\pi}$ and 
$f=\sqrt{\frac{6M^3c_1^2•}{•k}}$.  $^4 S_{(1)}$ is the contribution to the effective action from pure gravity and gravity coupled with the modular field $T(x)$. $^4 S_{(2)}$ is the kinetic term of the effective action and $^4 S_{(3)}$ is the contribution to the effective action purely from the modulus which in turn generates the modular potential. \\
                       
The modular potential  $V(\Phi)$ is  given by, 

\begin{eqnarray}
\nonumber V(\Phi)=[6k\omega ^4 ln \frac{\Phi}{f•} -3k\omega ^4 +2k\frac{\omega ^6}{c_2 ^2•}\frac{f^2}{\Phi ^2•}+2k\omega ^2 c_2 ^2 -\frac{1}{2•}k\frac{\omega ^8}{c_2 ^4•}-\frac{1}{2•}kc_2 ^4 \frac{\Phi ^4}{f^4•}\\ \nonumber
-3k\omega ^4(\frac{\omega ^2}{c_2 ^2• - \omega ^2}+\frac{c_2 ^2 \Phi ^2}{ \omega ^2 f^2-c_2 ^2• \Phi ^2 +})+\frac{2k \omega  ^6}{ c_2 ^2•-\omega ^2 }(\frac{\omega ^2}{c_2 ^2•}+\frac{c_2^2}{\omega ^2•})+\frac{2kc_2 ^2 \omega ^4 \Phi ^2}{ \omega ^2 f^2-c_2 ^2 \Phi ^2•}(\frac{\omega^2 f^2}{c_2 ^2 \Phi ^2•} + \frac{c_2 ^2 \Phi ^2••}{•\omega^2 f^2})
\\ -\frac{1}{2•}\frac{k \omega ^6}{c_2 ^2-\omega ^2•}(\frac{\omega ^4}{c_2 ^4•}+\frac{c_2 ^4}{•\omega ^4})-\frac{1}{2•}\frac{•k c_2 ^2 \omega^4 \Phi ^2}{\omega^2 f^2-c_2 ^2• \Phi ^2}(\frac{\omega^4 f^4}{c_2 ^4•\Phi ^4}+\frac{c_2^4•\Phi ^4•}{•\omega^4 f^4})]
\end{eqnarray}

It may be seen that in the limit $\omega \rightarrow 0$, we retrieve the Gravity-radion action as derived in
Goldberger \& Wise \citep{GW2000}. Clearly, the presence of $\omega$ generates a potential term for the modular field
$\Phi$. 

\begin{figure}
 \centering
\includegraphics[height=4.5cm,width=7.5cm]{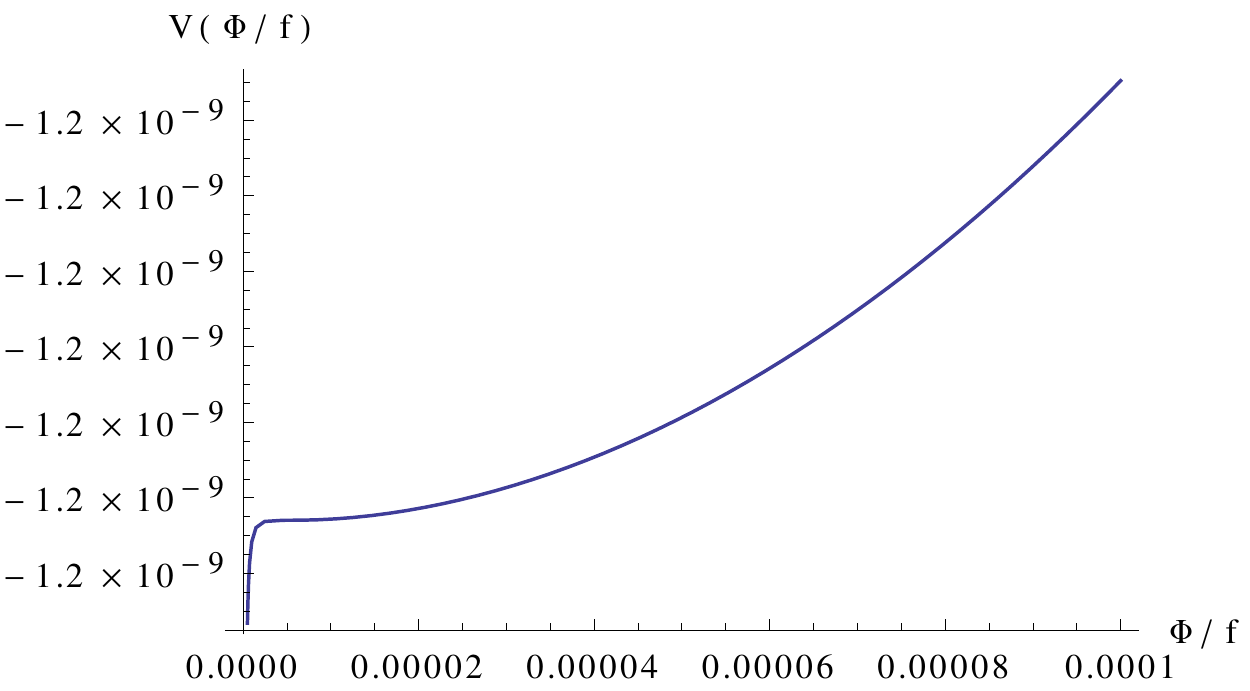}
\vskip-0.3cm
\caption{Modular potential for the DS brane for $\omega=10^{-5}$}
\label{lc}
\end{figure}

\begin{figure}
 \centering
\includegraphics[height=4.5cm,width=7.5cm]{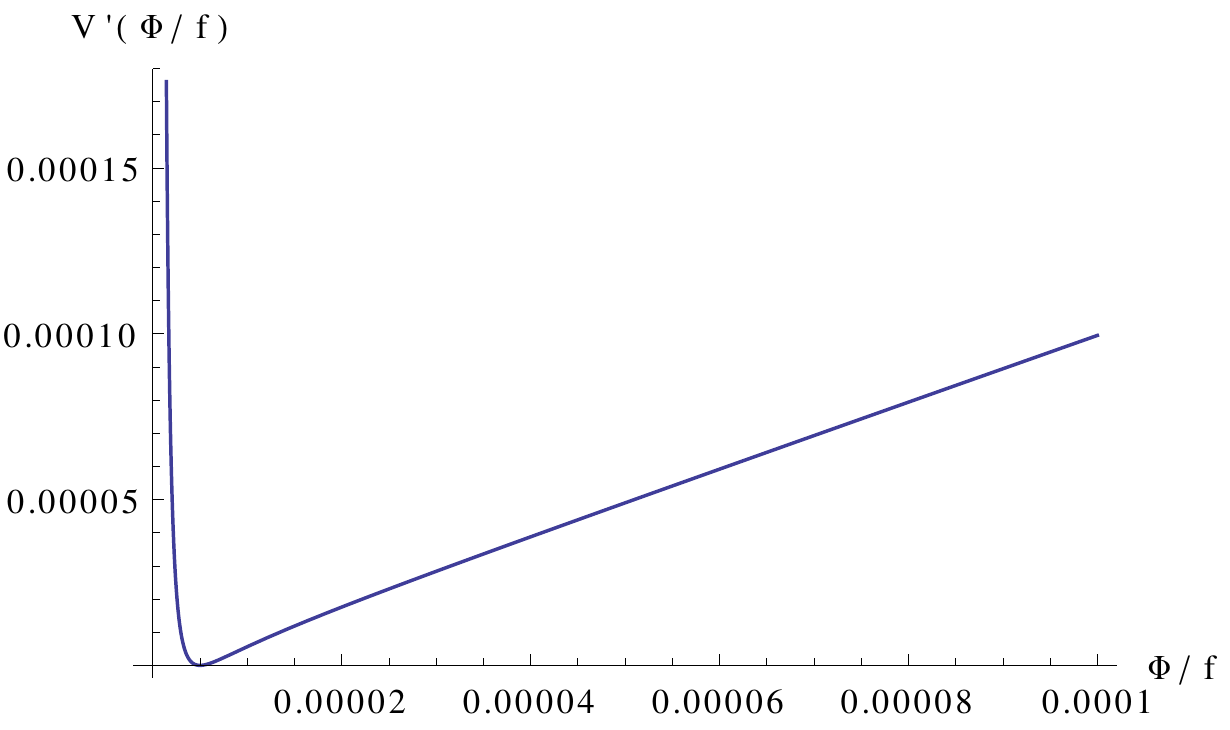}
\vskip-0.3cm
\caption{First derivative of the modular potential for the DS brane for $\omega=10^{-5}$}
\label{lc}
\end{figure}

To obtain the minimum of the potential, we find that its first derivative with respect to $\frac{\Phi}{f}$ is:


\begin{equation}
\frac{dV(\Phi/f)}{d(\Phi/f)•}=6M^3k\omega ^2 c_2 ^2[ \frac{\omega ^4 f^3}{c_2 ^4 \Phi^3•} +\frac{\Phi}{f•}-\frac{2\omega^2 f}{c_2^2 \Phi•}]
\end{equation}
From equation (28) it is clear that the first derivative of the potential vanishes when $\frac{\Phi}{f} \rightarrow \frac{\omega}{c_2•}$. 
The potential and its first derivative for $\omega=10^{-5}$ are illustrated in Figure 1 and Figure 2 respectively.
\\ \\ 

From Figure 1 and Figure 2 it is evident that the potential attains a metastable minima as $\frac{\Phi}{f}=e^{-kT(x)\pi}\rightarrow \frac{\omega}{c_2}$. One should note that, from the form of the warp factor given in equation (18), $\frac{\Phi}{f}< \frac{\omega}{c_2}$ is not possible. Hence, we consider, $\frac{\Phi}{f•}=\frac{\omega}{c_2}+\epsilon$ where, $\epsilon<<\frac{\omega}{c_2}$. Using this in equation (18) and 
neglecting higher order terms for $\epsilon$ we obtain
$e^{-A} \simeq c_2 \epsilon$. Adjusting the value of $\epsilon \sim 10^{-16}$ we find that  a proper resolution to the gauge hierarchy problem can be achieved.

\subsection*{•Case B:Anti-de Sitter 3 brane}
Here, we assume that the 3 branes are AdS branes. In such a scenario, the generalized warp factor
is given by:

\begin{equation}
e^{-A}=\omega cosh(ln \frac{\omega}{c_1•} + k|\phi|T(x)),
\end{equation}

Under such circumstances,

\begin{eqnarray}
\nonumber S_{gravity}^{(1)}=-2M^3\int d^4x \int_{-\pi}^{\pi} d\phi ~ \sqrt{-g} e^{-2A} [T(x)~ {^4}R + 8k^2e^{-2A} T(x) +12k^2e^{-2A} T(x)tanh^2 (ln \frac{•\omega}{c_1•}+kT(x)|\phi|) \\ \nonumber
-6k|\phi| T(x),^\alpha T(x),_\alpha tanh(ln \frac{•\omega}{•c_1} + kT(x)|\phi|)-6k^2|\phi|^2 T(x) T(x),^{\alpha}T(x),_{\alpha})tanh^2(ln \frac{•\omega}{•c_1} + kT(x)|\phi|)] 
\end{eqnarray}

and,

\begin{equation}
S_{gravity}^{(2)}= -\int d^4 x \int_{-\pi}^{\pi} e^{-4A} \sqrt {-g} ~ \Lambda
\end{equation}

Contribution to $S_{gravity}^{(1)}$ from the brane boundaries is the same as in the RS scenario.


The effective action for the AdS branes is:

\begin{eqnarray}
^4S = \int d^4 x (^4S_{(1)} +^4S_{(2)} +^4S_{(3)})
\end{eqnarray}

where,



    
where defining as before, $\Phi = fe^{-kT(x)\pi}$,

\begin{equation}
^4 S_{(1)}=-2M^3 \int d^4 x \sqrt {-g} ^4R[\frac{c_1 ^2}{4k}-\frac{\omega^2}{k•}ln\frac{\Phi}{f•}+\frac{\omega ^4}{4 k c_1^2•}\frac{f^2}{\Phi ^2•}-\frac{\omega^4}{4kc_1^2•}-\frac{c_1 ^2\Phi ^2}{4kf^2•}]
\end{equation} 
\begin{equation}
^4 S_{(2)}= \int d^4 x \sqrt{-g} [- \frac{1}{2•}\partial_\mu \Phi \partial^\mu  \Phi + 4 \frac{M^3}{k}\omega^2
(\frac{1}{\Phi ^2•}ln\frac{\Phi}{f•})\partial_\mu \Phi \partial^\mu  \Phi + \frac{3M^3}{•k}\frac{\omega^4}{c_1 ^2•}f^2 \frac{•\partial^\mu  \Phi \partial^\mu  \Phi}{•\Phi ^ 4}] 
\end{equation}

\begin{eqnarray}
^4 S_{3}=-2M^3 \int d^4 x \sqrt{-g} [V(\Phi)]
\end{eqnarray}

where, $V(\Phi)$ is the modular potential given by, 

\begin{eqnarray}
\nonumber V(\Phi)=[-6k\omega ^4 ln \frac{\Phi}{f•} -3k\omega ^4 -2k\frac{\omega ^6}{c_1 ^2•}\frac{f^2}{\Phi ^2•}-2k\omega ^2 c_1 ^2 -\frac{1}{2•}k\frac{\omega ^8}{c_1 ^4•}-\frac{1}{2•}kc_1 ^4 \frac{\Phi ^4}{f^4•}\\ \nonumber
3k\omega ^4(\frac{\omega ^2}{c_1 ^2• + \omega ^2}+\frac{c_1 ^2 \Phi ^2}{c_1 ^2• \Phi ^2 + \omega ^2 f^2})+\frac{2k \omega  ^6}{\omega ^2 + c_1 ^2•}(\frac{\omega ^2}{c_1 ^2•}+\frac{c_1 ^2}{\omega ^2•})+\frac{2kc_1 ^2 \omega ^4 \Phi ^2}{c_1 ^2 \Phi ^2 + \omega ^2 f^2•}(\frac{\omega^2 f^2}{c_1 ^2 \Phi ^2•} + \frac{c_1 ^2 \Phi ^2••}{•\omega^2 f^2})
\\ \frac{1}{2•}\frac{k \omega ^6}{\omega ^2• + c_1 ^2}(\frac{\omega ^4}{c_1 ^4•}+\frac{c_1 ^4}{•\omega ^4})+\frac{1}{2•}\frac{•k c_1 ^2 \omega^4 \Phi ^2}{c_1 ^2• \Phi ^2+\omega^2 f^2}(\frac{\omega^4 f^4}{c_1 ^4•\Phi ^4}+\frac{c_1 ^4•\Phi ^4•}{•\omega^4 f^4})]
\end{eqnarray}

To solve for the critical points of this potential, we find that its first derivative with respect to $\frac{\Phi}{f}$ is:

\begin{figure}
 \centering
\includegraphics[height=4.5cm,width=7.5cm]{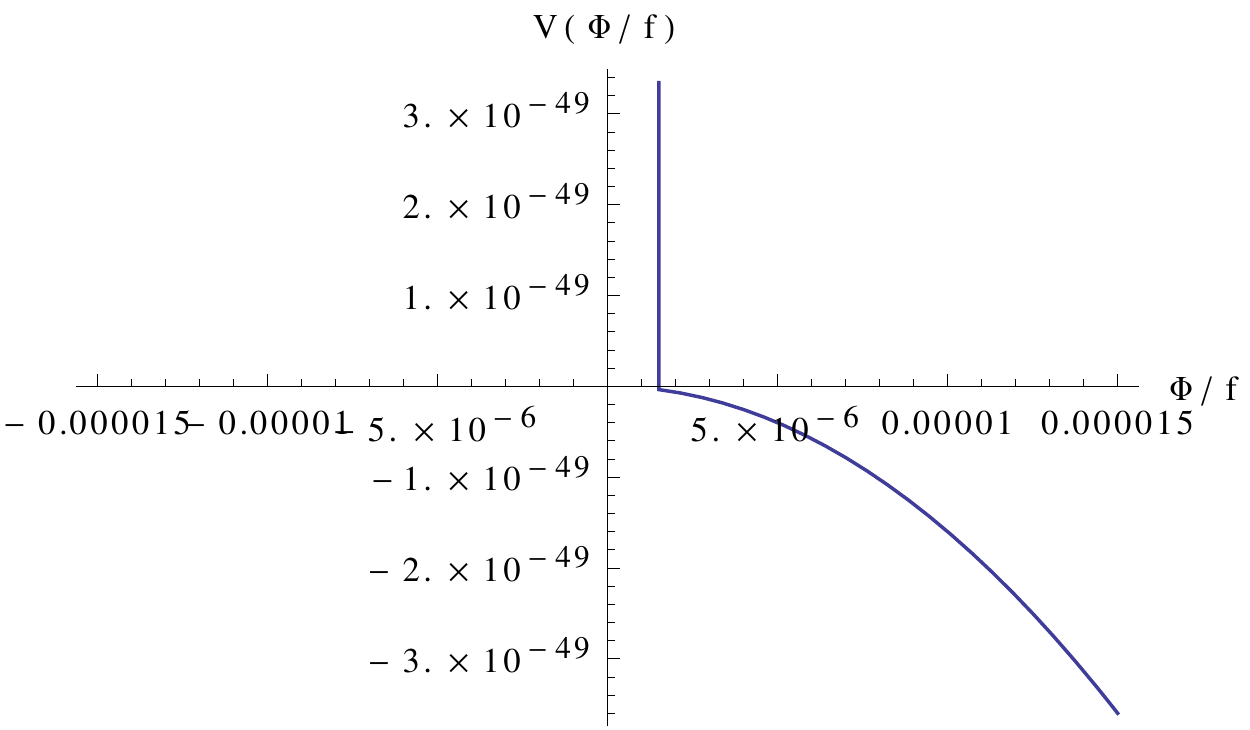}
\vskip-0.3cm
\caption{Modular potential for the ADS brane for $\omega=10^{-20}$}
\label{lc}
\end{figure}

\begin{figure}
 \centering
\includegraphics[height=4.5cm,width=7.5cm]{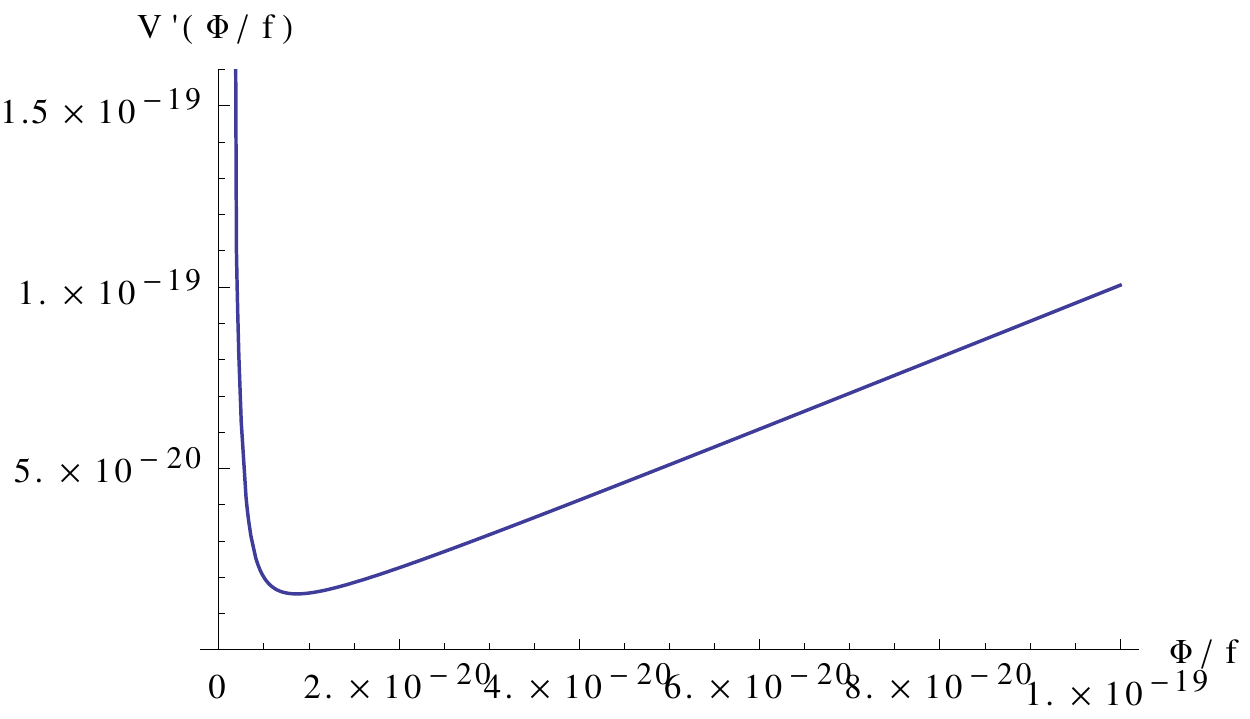}
\vskip-0.3cm
\caption{First derivative of the modular potential for the ADS brane for $\omega=10^{-20}$}
\label{lc}
\end{figure}

\begin{equation}
~or, \frac{dV(\Phi/f)}{d(\Phi/f)•}=6M^3k\omega ^2 c_1 ^2[ \frac{\omega ^4 f^3}{c_1 ^4 \Phi^3•} +\frac{\Phi}{f•}+\frac{2\omega^2 f}{c_1^2 \Phi•}]
\end{equation}
From equation (36) it is clear that the first derivative of the potential has no zero crossing. The potential does not have any minimum. The potential and its first derivative for $\omega=10^{-20}$ are shown in Figure 3 and Figure 4 respectively.
From the aforementioned figures it is quite evident that in the scenario of AdS three branes, the modular potential does not have any turning points. Thus, in this situation stability cannot be attained. However,the resolution to the gauge hierarchy problem is
independent of the radion field stabilization. Hence, in this scenario, the gauge hierarchy problem can be addressed as discussed in section 2, although the stability of the modular field cannot be achieved.

\section{Summary and Conclusion}
In this work, we explore the possibility of stabilizing the extra dimensional  modulus in the context of warped non-flat branes, without introducing any extra bulk scalar field. We show that the
a non-flat maximally symmetric brane automatically gives rise to a stable braneworld model when the brane cosmological constant is positive in otherwords it is de-Sitter in character.\\
In the current work, we include the dynamics of the modular field in the context of warped geometry in a curved 3-brane scenario that naturally incorporates the effect of the brane cosmological constant on modular stability. We find that the presence of the brane cosmological constant naturally generates a potential energy term for the modulus field in the Lagrangian of the effective action and no external scalar field is required   to stabilize the modulus.

We further show that if the branes are anti-de Sitter then the radion potential which arises self-consistently due to the presence of the brane vacuum energy, does not have any turning points (Figure 3 and Figure 4) and hence modular stabilization cannot be achieved in such a scenario. However, the resolution to the gauge hierarchy problem which is independent of modular stabilization can always be attained. This is discussed in \S3.
On the other hand, if the 3 branes are de Sitter branes, then the modular potential has a metastable minimum at which the radion field is stabilized ( See figure 1 and figure 2 )
The value of this minimum depends on the brane vacuum energy and by tuning the brane cosmological constant
appropriately the gauge hierarchy problem can be resolved at the minimum of this potential. Thus, the stabilization of the modular field and the resolution to the gauge hierarchy problem in the context of non-flat 3-branes in a warped brane world scenario can all be addressed concomitantly.
The fact that  we live presently in a de Sitter universe with a tiny cosmological constant therefore can account for the stability of modulus and points towards a stable braneworld description of our Universe.




\end{document}